\documentclass[11pt]{article}
\usepackage{psfrag}
\usepackage{amssymb}
\usepackage{graphicx}
\usepackage{url}

\newcounter{proclineno}

\makeatletter
\def\@currentproclabel{}

\def\proclabel#1{\let\@currentproclabel\theproclineno\@bsphack\if@filesw%
{\let\thepage\relax\def\protect{\noexpand\noexpand\noexpand}%
\edef\@tempa{\write\@auxout{\string\newlabel{#1}{{\@currentproclabel}{\thepage}}}}%
\expandafter}\@tempa\if@nobreak \ifvmode\nobreak\fi\fi\fi\@esphack}%
\makeatother
\newtheorem{THEOREM}{Theorem}
\newenvironment{theorem}{\begin{THEOREM} \hspace{-.85em} {\bf .} }%
                        {\end{THEOREM}}
\newtheorem{LEMMA}{Lemma}
\newenvironment{lemma}{\begin{LEMMA} \hspace{-.85em} {\bf .} }%
                      {\end{LEMMA}}
\newtheorem{COROLLARY}{Corollary}
\newenvironment{corollary}{\begin{COROLLARY} \hspace{-.85em} {\bf .} }%
                          {\end{COROLLARY}}
\newtheorem{PROPOSITION}{Proposition}
\newenvironment{proposition}{\begin{PROPOSITION} \hspace{-.85em} {\bf .} }%
                            {\end{PROPOSITION}}
\newtheorem{PROPERTY}{\normalfont\textit{Property}}
\newenvironment{property}{\begin{PROPERTY} \hspace{-.85em} {\itshape .} \rm}%
                            {\end{PROPERTY}}
\newtheorem{DEFINITION}{Definition}
\newenvironment{definition}{\begin{DEFINITION} \hspace{-.85em} {\bf .} \rm}%
                            {\end{DEFINITION}}
\newtheorem{DEFINITIONL}[DEFINITION]{Definition}
                            {\end{DEFINITIONL}}
\newtheorem{EXAMPLE}{Example}
                            {\end{EXAMPLE}}
\newtheorem{ALGORITHM}{Algorithm}
                          {\end{ALGORITHM}}
\newtheorem{PROCEDURE}{Procedure}
                          {\end{PROCEDURE}}
\newtheorem{REMARK}{Remark}
                            {\end{REMARK}}
\newtheorem{CLAIM}{Claim}
                            {\end{CLAIM}}
\newtheorem{CLAIMEMPH}{Claim}
                            {\end{CLAIMEMPH}}
\newtheorem{HYPOTHESIS}{Hypothesis}
                            {\end{HYPOTHESIS}}
\newtheorem{HYPOTHESISL}[HYPOTHESIS]{Hypothesis}
                            {\end{HYPOTHESISL}}
\newtheorem{FACT}{Fact}
                            {\end{FACT}}
\newtheorem{FACTL}[FACT]{Fact}
                            {\end{FACTL}}

\newcommand{\prf}{\noindent{\bf Proof.} }
\newcommand{\eprf}{}

\newcommand{\qed}{\hfill \ensuremath{\Box}}

\begin{document}


\title{Minimum regulation of uncoordinated matchings}
\date{}


 \author{Bruno Escoffier$^{1,2}$ \and Laurent Gourv\`es$^{2,1}$ \and J\'er\^ome
 Monnot$^{2,1}$\\
{\small 1. Universit\'e de Paris-Dauphine, LAMSADE, 75775 Paris, France}\\
{\small 2.  CNRS, FRE 3234, 75775 Paris, France} \\
{\small $\{$escoffier, laurent.gourves,
monnot$\}$@lamsade.dauphine.fr}}

%


\maketitle

\begin{abstract} Due to the lack of coordination, it is unlikely that the selfish players of a strategic game reach a socially good state.
A possible way to cope with selfishness is to compute a desired outcome (if it is tractable) and
impose it. However this answer is often inappropriate 
because compelling an agent can be costly, unpopular or just hard to
implement. Since both situations (no coordination and full
coordination) show opposite advantages and drawbacks, it is natural
to study possible tradeoffs.
In this paper we study a strategic game where the nodes of a simple graph $G$ are independent agents who try to form pairs: e.g. jobs and applicants, tennis players for a  match, etc. In many instances of the game, a Nash equilibrium significantly deviates from a social optimum. 
We analyze a scenario where we fix the strategy of some players; the other players are free to make their choice. The goal is to compel a minimum number of players and guarantee that any possible equilibrium of the modified game is a social optimum, i.e. created pairs must form a maximum matching of $G$. We mainly show that this intriguing problem is NP-hard and propose an approximation algorithm with a constant ratio.


\end{abstract}

\section{Introduction}

We propose to analyze the following non cooperative game. The input is a simple graph $G=(V,E)$ where every vertex is
controlled by a player whose strategy set is his neighborhood in $G$.
If a vertex $v$ selects a neighbor $u$ while $u$ selects $v$ then the two nodes are matched and they both have utility $1$. If a
vertex $v$ selects a neighbor $u$ but $u$ does not select $v$ then
$v$ is unmatched and its utility is $0$.

Matchings in graphs are a model for many practical situations where nodes may represent autonomous entities. For instance,
suppose that each node is a chess (or tennis) player searching for a partner. An edge between two players means that
they  are available at the same time, or just that they know each other. As another example, consider a set of companies on one side, each offering a job, and on the other side a set of applicants. There is an edge if the worker is qualified for the job.


Taking the number of matched nodes as the social welfare associated
with a strategy profile (a maximum cardinality matching is then  a social optimum),
we can rapidly observe that the game has a
high {\em price of anarchy}. The system needs regulation because the uncoordinated and selfish
behavior of the players deteriorates its performance. How can we do this regulation? One can compute a
maximum matching (in polynomial time) and force the players to follow
it. However forcing some nodes' strategy may be costly, unpopular or simply hard to implement.
When both cases (complete freedom and total regulation) are not satisfactory, it is necessary to make a tradeoff.
In this paper we propose to fix the strategy of some nodes; the other players are free to make their choice.
The only requirement is that every equilibrium of the modified game is a social optimum (a maximum matching).
Because it is unpopular/costly, the number of forced players should be {\em minimum}.
We call the optimization problem {\sc mfv} for {\em minimum forced vertices}. The challenging task is to identify the nodes
which play a central role in the graph.

\subsection{Related work}

There is a great interest in how uncoordinated and selfish agents make use of a common resource \cite{RT02,KP09}. A popular way of modeling the problem is by means of a noncooperative
game and by viewing its equilibria as outcomes of selfish behavior. In this context,
the {\em price of anarchy} (PoA) \cite{KP09}, defined as  the value of the worst Nash equilibrium relative
to the social optimum, is a well established measure of the performance
deterioration. A game with a high PoA needs regulation and several ways to improve the system performance
exist in the literature, including {\em coordination mechanisms} \cite{CKN09,ILMS05} and {\em Stackelberg strategies} \cite{RT04,CS07,BHS10}.


In \cite{RT04} T. Roughgarden studies a nonatomic scheduling problem where a rate of flow $r$ should be to assigned to a set of parallel machines
with load dependent latencies. There are two kinds of players: a leader controlling a fraction $\alpha$ of $r$
and a set of followers, everyone handles an infinitesimal part of $(1-\alpha)r$. The leader, interested in optimizing the total latency,
plays first (i.e. assigns $\alpha r$ to the machines) and keeps his strategy fixed.
The followers react independently and selfishly to the leader's strategy, optimizing their own latency.
The author gives an algorithm for computing a leader strategy that induces an equilibrium with total latency
no more than $1/\alpha$ times that of the optimal assignment of jobs to machines.  
He also mentions that his approach falls into the area of {\em Stackelberg games} \cite{RT04}. 
The {\sc mfv} problem introduced and studied in this article follows a rather similar approach. Instead of optimizing the social welfare with a given rate of control on the players, one tries to minimize the control on the players while an optimal social welfare is guaranteed.  In other words a leader interested in the social welfare fixes the strategy of a minimum number of nodes so that any equilibrium reached by the unforced nodes creates a maximum number of pairs. 

The {\sc mfv} problem is related to the well known {\em stable marriage problem} ({\sc smp}) \cite{GS62}. In the {\sc smp} there are $n$ women and $n$ men who rank the persons of the opposite sex in a strict order of preference. A solution is a matching of size $n$; it is {\em unstable} if two participants 
prefer being together than being with their respective partner. Interestingly a stable matching always exists and one can compute it with the algorithm of Gale and Shapley \cite{GS62}. Many variants of the {\sc smp} were studied in the literature: all participants are of the same gender (the {\em stable roommates problem}) \cite{Irving85}, ties in preferences are allowed \cite{IMMM99}, players can give an incomplete list \cite{IMMM99}, etc. In fact 
the {\sc mfv} problem has some similarities with 
the stable roommates problem with simplified preferences: every participant gives a list of equivalent/interchangeable partners, omitting only those persons he would never
accept under any circumstances.



\subsection{Our results}

We first give a formal definition of the noncooperative game and
show that it has a high price of anarchy. An associated optimization
problem ({\sc mfv}) is then introduced. In Section \ref{secfeasible}
we show that we can decide in polynomial time whether a solution is
feasible or not. In particular one can detect graphs for which any
pure Nash equilibrium corresponds to a maximum matching though no
vertex is forced.

Next we investigate the complexity and the approximability of {\sc
mfv}. Interestingly the problem in graphs admitting a perfect
matching is equivalent to the vertex cover problem (see Subsection
\ref{subsecperfect}). In Subsection \ref{subsecgeneral} we propose a
$6$-approximation algorithm called \texttt{APPROX} for general
graphs. A part of the proof showing that \texttt{APPROX} is a
$6$-approximation is given in Section \ref{sectech}. Concluding
remarks are given in Section \ref{secconcl}. 

\section{The strategic game and the optimization problem}
We are given a simple connected graph $G=(V,E)$. Every  vertex is
controlled by a player so we interchangeably mention a vertex and
the player who controls it. The strategy set of every player $i$ is
his neighborhood in $G$, denoted by $\mathcal{N}_G(i)$. Then the
strategy set of a leaf in $G$ is a singleton. Throughout the article
$S_i$ designates the action/strategy of player $i$. A player is
matched if the neighbor that he selects also selects him. Then $i$
is matched under $S$ if $S_{S_i}=i$. A player has {\em utility} $1$
when he is matched, otherwise it is $0$. The utility of player $i$
under state $S$ is denoted by $u_i(S)$.

The {\em social welfare} is defined as the number of matched nodes. 
We focus on the pure strategy Nash equilibria, considering them as
the possible outcomes of the game. It is not difficult to see that
every instance admits a pure Nash equilibrium. In addition, the
players converge to a Nash equilibrium after at most $|V|/2$ rounds.

Interestingly there are some graphs for which the players always
reach a social optimum: paths of length 1, 2 and 4; cycles of length
3 and 5;  stars, etc. However the social welfare can be very far
from the social optimum in many instances as the following result
states.


\begin{theorem} \label{theopoa} The PoA is $\max \{2/ |V|, 1/ \Delta\}$ where $\Delta$ denotes the maximum degree of a node.
\end{theorem}

\prf The {\em social welfare} is  denoted by $\mathcal{SW}(S)$ under state $S$. Let $S$ and $S^*$ be a Nash equilibrium and an optimum state respectively. 
First remark that at least two players are matched in $S$, i.e.
$\mathcal{SW}(S) \ge 2$. Indeed, take any player $i$. If $S_{S_i}=i$
then $i$ and $S_i$ are matched. If $S_{S_i} \neq i$ then $S_i$ must
be matched with a node $j \neq i$ because $S$ is a Nash equilibrium.
Using the fact that $\mathcal{SW}(S^*) \le |V|$, it follows that PoA
$\leq 2/|V|$. A complete graph gives a tight example.

Now let us show that PoA $\leq 1/ \Delta$. Since $u_{S_i}(S)$ must
be equal to 1 for every $i \in V$, otherwise $S$ is not a Nash
equilibrium, we get that $\max_{j \in \mathcal{N}_G(i)} u_j(S) \ge
1$ for every $i \in V$. We deduce that
$$\sum_{i \in V} \sum_{j \in \mathcal{N}_G(i)} u_j(S) \ge \sum_{i \in V} \max_{j \in \mathcal{N}_G(i)} u_j(S) \ge |V| \ge \mathcal{SW} (S^*).$$
We also remark that
$$\sum_{i \in V} \sum_{j \in \mathcal{N}_G(i)} u_j(S) \le \Delta \sum_{i \in V} u_i(S) = \Delta \mathcal{SW}(S).$$
Combining the previous two inequalities, we get that
$\mathcal{SW}(S) / \mathcal{SW}(S^*) \ge 1/ \Delta$. A tight example
for any given $\Delta$ can be the following:
\begin{itemize}
    \item the vertex set is made of $2\Delta$ nodes denoted by $\{x,y\} \cup \{v_1, \cdots,v_{\Delta -1}\} \cup \{v'_1, \cdots,v'_{\Delta -1}\}$
  \item for every $i \in [1.. \Delta -1]$ build the edges $(v_i,v'_i)$, $(x,v_i)$ and $(y,v'_i)$.
  \item add the edge $(x,y)$.
\end{itemize}
Clearly both $x$ and $y$ have degree $\Delta$. The state $S$ where
$S_{v_i}=x$, $S_{v'_i}=y$, $S_{x}=y$ and $S_{y}=x$  is a Nash
equilibrium of social welfare $2$ (only $x$ and $y$ are matched).
The state $S^*$ where $v_i$ is matched with $v'_i$ for every $i$,
and $x$ is matched with $y$, is a an optimum state of value $2
\Delta$.  \qed

\eprf

Theorem~\ref{theopoa} indicates that the system needs regulation to
achieve an acceptable state where a maximum number of players are
matched. That is why we introduce a related optimization problem,
called {\sc mfv} for {\em minimum forced vertices}.

For a graph $G=(V,E)$, instance of the {\sc mfv} problem,
a solution is a pair $\langle T, Q \rangle$ where 
$Q$ is a subset of players and every player in $Q$ is forced to
select node $T_i \in \mathcal{N}_G(i)$ (i.e. $T=(T_i)_{i\in Q}$). In
the following $\langle T, Q \rangle$ is called a {\em Stackelberg
strategy} or simply a solution. A state $S$ is a {\em Stackelberg
equilibrium} resulting from the Stackelberg strategy $\langle T, Q
\rangle$ if $S_i=T_i$ for every $i \in Q$, and $\forall i \in V
\setminus Q$, $\forall j \in \mathcal{N}_G(i)$, $u_i(S) \ge
u_i(S_{-i},j)$. Here $(S_{-i},j)$ denotes $S$ where $S_i$ is set to
$j$.

A solution $\langle T, Q \rangle$ to the {\sc mfv} problem is said feasible if every Stackelberg equilibrium is a social optimum. The value of
$\langle T, Q \rangle$ is  $|Q|$ to be minimized.

Now let us introduce some notions that we use throughout the article.
The matching induced by a strategy profile $S$  is denoted by ${\cal
M}^S$ and defined as $\{ (u,v) \in E : S_u=v \textrm{ and }
S_v=u\}$. We also define three useful notions of compatibility:
\begin{itemize}
    \item A matching $M$ and a state $S$ are compatible if $M = {\cal M}^S$ 
\item  A state $S$ and a solution $\langle T,Q \rangle$ are compatible if $T_i=S_i$ for all $i \in Q$
\item
A matching $M$ and a solution $\langle T,Q \rangle$ are compatible if there exists a state  $S$ compatible with both $M$ and $\langle T,Q\rangle$.
\end{itemize}
We sometimes write that a matching is {\em induced} by a solution if they are compatible.












\section{Feasible solutions of the {\sc mfv} problem} \label{secfeasible}

Though it is easy to produce a feasible solution to any instance of
the {\sc mfv} problem, deciding whether a given solution, even the
empty one, is feasible is not straightforward.

A maximum matching compatible with a solution $<T,Q>$ can be computed in polynomial time: start from $G$, remove every edge
$(u,v)$ such that $u$ is forced to play a node $w$ different from $v$, and compute a maximum matching in the resulting graph.
If the result is not a maximum matching in $G$ then it is clear that $\langle T,Q \rangle$ is not feasible.
However it is not necessarily feasible when the resulting matching  is maximum in $G$. 
In the sequel, $\mathcal{M}^*$ denotes a matching compatible with $<T,Q>$ and we assume that $\mathcal{M}^*$ is maximum in $G$.
In addition $S^*$ denotes a Stackelberg equilibrium compatible with $<T,Q>$ and $\mathcal{M}^*$.

We resort to patterns called {\em diminishing configurations}. 

\begin{definition} \begin{itemize} \item $\mathcal{M}^*$ and $\langle T,Q \rangle$ possess a long diminishing configuration if there are
$2r$ vertices $v_1, \ldots, v_{2r}$ arranged in a path as on Figure \ref{figcas13} $(a)$ ($r$ is an integer such that $r \ge 2$) and a strategy profile $S^*$ satisfying
\begin{itemize}
\item $S^*$ is a Stackelberg equilibrium compatible with $<T,Q>$ and $\mathcal{M}^*$
\item if $v_1 \notin Q$ then there is no $v \in V \setminus \{v_1, \ldots, v_{2r}\}$ such that $S^*_v=v_1$
\item if $v_{2r} \notin Q$ then there is no $v \in V \setminus \{v_1, \ldots, v_{2r}\}$ such that $S^*_v=v_{2r}$
\end{itemize}

\begin{figure}
\begin{center}
\psfrag{a}{$v_1$}
\psfrag{b}{$v_2$}
\psfrag{c}{$v_3$}
\psfrag{d}{$v_{2r-1}$}
\psfrag{e}{$v_{2r}$}
\psfrag{f}{$x$}
\psfrag{g}{$y$}
\psfrag{h}{$z$}
\psfrag{x}{forced node}
\psfrag{y}{either forced or unforced node}
\psfrag{z}{unforced node}
\psfrag{c1}{$(a)$}
\psfrag{c2}{$(b)$}
\psfrag{c3}{$(c)$}
\psfrag{c4}{$(d)$}
\psfrag{c5}{$(e)$}
\psfrag{c6}{$(f)$}
\psfrag{c7}{$(g)$}
\psfrag{c8}{$(h)$}
\includegraphics[width=90mm]{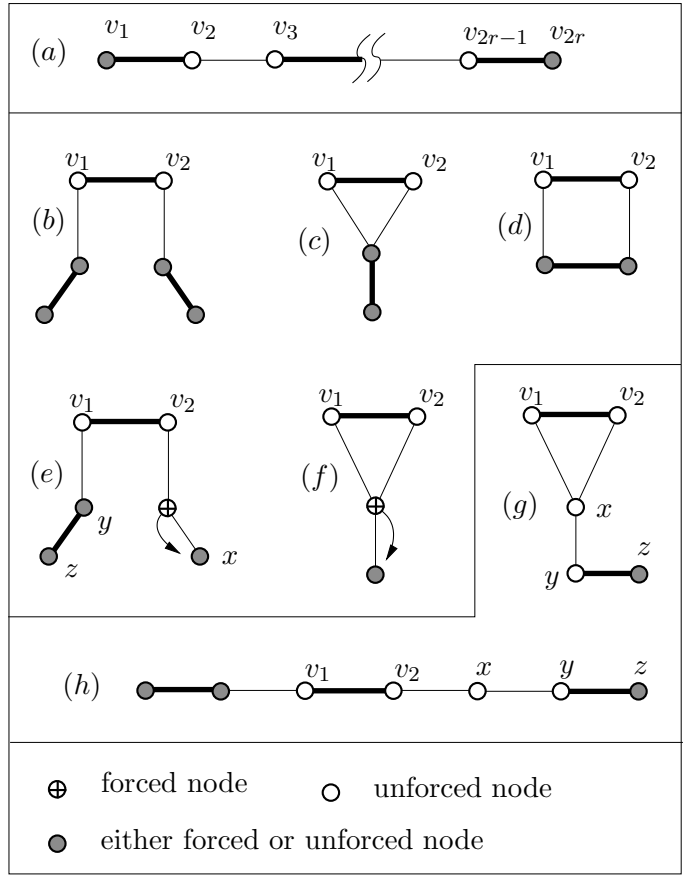}
\caption{The eight diminishing configurations. Every bold edge belongs to $\mathcal{M}^*$, every thin edge belongs to
$E \setminus \mathcal{M}^*$. A white node is not in $Q$ while crossed node must belong to $Q$. Grey nodes can be in $Q$ or not. For the case $(e)$, nodes $x$ and $y$ (resp. $z$) can be the same.} \label{figcas13}
\end{center}
\end{figure}


\item $\mathcal{M}^*$ and $\langle T,Q \rangle$ possess a short diminishing configuration
if there exists one pattern among those depicted on Figure \ref{figcas13} $(b)$ to $(f)$
and a strategy profile $S^*$ satisfying
\begin{itemize}
    \item $S^*$ is a Stackelberg equilibrium compatible with $<T,Q>$ and $\mathcal{M}^*$
    \item there is no node $v \in V \setminus \{v_1, v_{2}\}$  such that $S^*_v \in \{v_1,v_{2}\}$
\end{itemize}


\item $\mathcal{M}^*$ and  $\langle T,Q \rangle$ possess an average diminishing configuration
if there exists one pattern among those depicted on Figure \ref{figcas13} $(g)$ and $(h)$ and a strategy profile $S^*$ satisfying
\begin{itemize}
    \item $S^*$ is a Stackelberg equilibrium compatible with $<T,Q>$ and $\mathcal{M}^*$
    \item there is no $v \in V \setminus \{v_1,v_2,y\}$ such that $S^*_v \in \{v_1,v_{2},z\}$
\end{itemize}
\end{itemize}

\end{definition}

In the following Lemma, we assume that $\mathcal{M}^*$, the maximum matching compatible with $\langle T,Q\rangle$, is also maximum in $G$.

\begin{lemma} \label{car} $\langle T,Q\rangle$ is not feasible if and only if $\mathcal{M}^*$ and $\langle T,Q\rangle$ possess a diminishing configuration.
\end{lemma}

\prf

\begin{figure}
\begin{center}
\psfrag{a}{$v_1$} \psfrag{b}{$v_2$} \psfrag{c}{$v_3$}
\psfrag{d}{$v_{2r-1}$} \psfrag{e}{$v_{2r}$} \psfrag{f}{$x$}
\psfrag{g}{$y$} \psfrag{h}{$z$} \psfrag{c1}{$(a)$}
\psfrag{c2}{$(b)$} \psfrag{c3}{$(c)$} \psfrag{c4}{$(d)$}
\psfrag{c5}{$(e)$} \psfrag{c6}{$(f)$} \psfrag{c7}{$(g)$}
\psfrag{c8}{$(h)$}
\includegraphics[width=90mm]{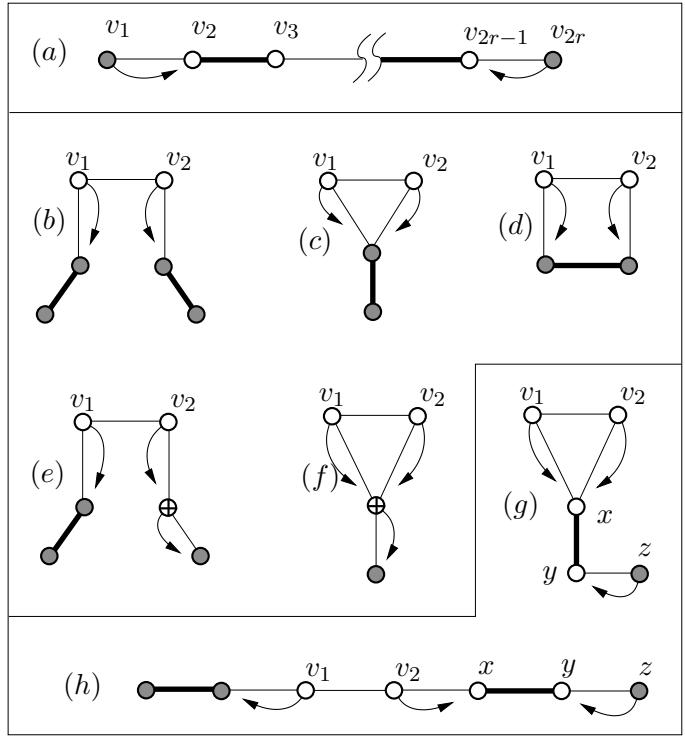}
\caption{For each configuration of Figure \ref{figcas13} there is a
way to decrease the matching by one unit. The corresponding strategy
profile remains a Nash equilibrium compatible with $\langle
T,Q\rangle$.} \label{figcas21}
\end{center}
\end{figure}

$(\Leftarrow)$ Suppose that $\mathcal{M}^*$ and $\langle T,Q\rangle$
possess a diminishing configuration. One can slightly modify $S^*$,
as done on Figure \ref{figcas21} for each case, such that the
strategy profile remains a Stackelberg equilibrium and the
corresponding matching has decreased by one unit. Therefore $\langle
T,Q\rangle$ is not feasible.

\medskip

\begin{figure}
\begin{center}
\includegraphics[width=60mm]{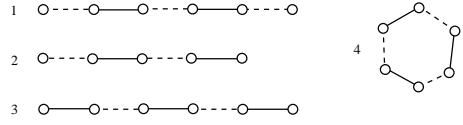}
\caption{The four cases for the connected components of
$\mathcal{M}' \Delta \mathcal{M}^*$. Edges of $\mathcal{M}^*$ and
$\mathcal{M}'$ are respectively solid and dashed.} \label{figcompo}
\end{center}
\end{figure}

\noindent $(\Rightarrow)$ Let $S'$ be a Stackelberg equilibrium
compatible with $\langle T,Q\rangle$ such that its associated
matching $\mathcal{M}'$ is not maximum in $G$. Consider the
symmetric difference $\mathcal{M}' \Delta \mathcal{M}^*$. Its
connected components are of four kinds:
\begin{itemize}
    \item a path which starts with an edge of $\mathcal{M}'$, alternates edges of $\mathcal{M}^*$ and $\mathcal{M}'$, and ends with an edge of
      $\mathcal{M}'$ (see case 1 in Figure \ref{figcompo})
    \item a path which starts with an edge of $\mathcal{M}'$, alternates edges of $\mathcal{M}^*$ and $\mathcal{M}'$, and ends with an edge of
      $\mathcal{M}^*$  (see case 2 in Figure \ref{figcompo})
    \item a path which starts with an edge of $\mathcal{M}^*$, alternates edges of $\mathcal{M}^*$ and $\mathcal{M}'$, and ends with an edge of
      $\mathcal{M}^*$  (see case 3 in Figure \ref{figcompo})
    \item an even cycle which alternates edges of $\mathcal{M}^*$ and $\mathcal{M}'$  (see case 4 in Figure \ref{figcompo})
\end{itemize}

Since  $|\mathcal{M}^*| > |\mathcal{M}'|$ there must be one
component of the third kind because this is the only case which
contains more edges of $\mathcal{M}^*$ than edges of $\mathcal{M}'$.
Notice that the two nodes on the extremities of this component are
unmatched in $\mathcal{M}'$. We consider two cases, whether this
component contains at least two edges of $\mathcal{M}^*$ (Case A),
or just one (Case B).

\medskip

\noindent Case A: Let us denote by $v_1, \ldots, v_{2r}$ the nodes
of the component. Since $S'_v \neq S^*_v$ holds for every $v \in
\{v_2, \cdots, v_{2r-1}\}$, we deduce that $\{v_2, \cdots,
v_{2r-1}\} \cap Q = \emptyset$. If $v_1$ and $v_{2r}$ belong to $Q$
then $\{v_1, \cdots, v_{2r}\}$ and $S^*$ form a long diminishing
configuration. Suppose that neither $v_1$ nor $v_{2r}$ belong to
$Q$. If there is no vertex $v \in V \setminus \{v_1, \ldots,
v_{2r}\}$ such that $S^*_v \in \{ v_1,v_{2r} \}$ then $\{v_1,
\cdots, v_{2r}\}$ and $S^*$ form a long diminishing configuration.
If there is a node $v \in Q$ such that $T_v \in \{v_1,v_{2r}\}$ then
$S'$ is not a Stackelberg equilibrium ($v_1$ and $v_{2r}$ are
unforced and unmatched in $\mathcal{M}'$ so one of them can play $v$
and be matched), contradiction. If there is a node $v \notin Q \cup
\{v_2, \cdots, v_{2r-1}\}$ such that $S^*_v \in \{v_1,v_{2r}\}$ then
$v$ is unmatched in $\mathcal{M}^*$, all its neighbors are matched
because $\mathcal{M}^*$ is maximum, and $S'_v \notin \{v_1,v_{2r}\}$
because $S'$ is a Stackelberg equilibrium. One can set $S^*_v
\leftarrow S'_v$ every time this case happens and deduce that
$\{v_1, \cdots, v_{2r}\}$ and $S^*$ form a long diminishing
configuration. Indeed $S^*$ though modified remains a Stackelberg
equilibrium compatible with $\mathcal{M}^*$ and $\langle
T,Q\rangle$.

The last case is when $v_1 \notin Q$ while $v_{2r} \in Q$ (the case
$v_{2r} \notin Q$ while $v_1 \in Q$ is completely symetric). If
there is no node $v \in V \setminus \{v_1, \cdots,v_{2r}\}$ such
that $S^*_v = v_1$ then $\{v_1, \cdots, v_{2r}\}$ and $S^*$ form a
long diminishing configuration. If there is a node $v \in Q$ such
that $T_v = v_1$ then $S'$ is not a Stackelberg equilibrium,
contradiction. If there is a node $v \notin Q$ such that $S^*_v =
v_1$ then $S'_v \neq v_1$ because $S'$ is a Stackelberg equilibrium.
One can set $S^*_v \leftarrow S'_v$ every time this case happens and
deduce that $\{v_1, \cdots, v_{2r}\}$ and $S^*$ form a long
diminishing configuration.

\medskip

\begin{figure}
\begin{center}
\psfrag{a}{$v_1$} \psfrag{b}{$v_2$} \psfrag{c}{$v_3$}
\psfrag{d}{$v_{4}$} \psfrag{e}{$v_{5}$} \psfrag{f}{$v_{6}$}
\psfrag{g}{$v_{7}$} \psfrag{1}{$(B1)$} \psfrag{2}{$(B2)$}
\psfrag{3}{$(B3)$} \psfrag{4}{$(B4)$} \psfrag{5}{$(B5)$}
\includegraphics[width=118mm]{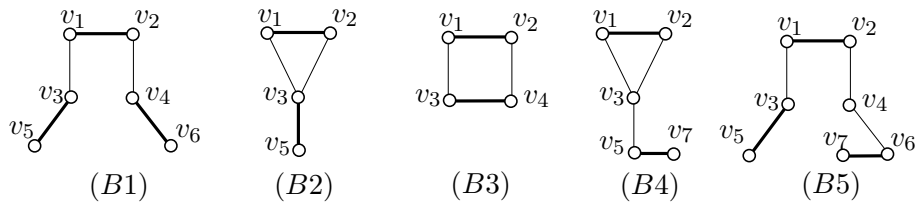}
\caption{Bold edges belong to $\mathcal{M}^*$ and thin edges exist
in the graph.} \label{figcas4}
\end{center}
\end{figure}

\noindent Case B: Let $v_1$ and $v_2$ be the two nodes of the
component. These nodes are matched together in $\mathcal{M}^*$ but
unmatched in $\mathcal{M}'$ so $\{v_1,v_2\} \cap Q = \emptyset$,
$S'_{v_1} \neq v_2$ and $S'_{v_2} \neq v_1$ (because $S'$ is a
Stackelberg equilibrium). Let us suppose that $S'_{v_1}=v_3$ and
$S'_{v_2}=v_4$. It is possible that $v_3=v_4$. Since $S'$ is a
Stackelberg equilibrium, $S'_{v_3} \notin \{v_1,v_2\}$ and $S'_{v_4}
\notin \{v_1,v_2\}$. We assume that $S'_{v_3}=v_5$ and
$S'_{v_4}=v_6$. Of course $v_5=v_6$ when $v_3=v_4$.  It is possible
that $v_3=v_6$ and $v_4=v_5$ when $v_3 \neq v_4$. If $v_3$ and $v_4$
are both unmatched in $\mathcal{M}^*$ (and $v_3 \neq v_4$) then
$(v_3,v_1,v_2,v_4)$ is an augmenting path,  contradicting the
optimality of $\mathcal{M}^*$. Then we can list 5 different cases,
denoted by $B1$ to $B5$, and depicted on Figure \ref{figcas4}. For
case $B4$ (resp. $B5$), $v_5$ (resp. $v_6$) must be matched with a
node that we denote by $v_7$, since otherwise $\mathcal{M}^*$ is not
optimal.

Before we analyze the 5 cases, we focus on the neighbors of $v_1$ or
$v_2$ which are unmatched in $\mathcal{M}^*$:
\begin{itemize}
\item Suppose there is a node $v$ such that $v \notin Q$, $S^*_{v}=v_1$, $v \neq v_2$ (resp. $S^*_{v}=v_2$, $v \neq v_1$) and $v$ has a neighbor $w \notin \{v_1,v_2\}$. Node $v$ is unmatched in $\mathcal{M}^*$ but $w$ is matched because $\mathcal{M}^*$ is maximum. Then modify $S^*$ and set $S^*_{v}=w$. 
 This modification is done each time it is possible. Remark that $S^*$ remains a Stackelberg equilibrium    after the modification.
    \item Suppose there is a node $v$ such that $v \notin Q$ and $v$ has no neighbor $w \notin \{v_1,v_2\}$. Then $\mathcal{N}_G(v)  \subseteq  \{v_1 , v_2\}$ and it contradicts the fact that $S'$ is a Stackelberg equilibrium because $v_1$ or $v_2$ could change its strategy (play $v$) and benefit.
    \item Suppose there is a node $v$ such that $v \in Q$ and
$T_v \in \{v_1,v_2\}$, i.e. $v$ is forced to play $v_1$ or $v_2$. It
contradicts the fact that $S'$ is a Stackelberg equilibrium because
$v_1$ or $v_2$ could change its strategy, play $v$, and benefit.
\end{itemize}

It follows that we can assume that there is no node $v$ such that
$S^*_v \in \{v_1,v_2\}$.

\medskip

\noindent Cases $B1$, $B2$ and $B3$: These cases correspond to the
short diminishing configurations of Figure \ref{figcas13} $(b)$,
$(c)$ and $(d)$ respectively.

\medskip

\noindent Case $B4$: 
Suppose that $v_3 \in Q$. If $T_{v_3} \in \{v_1,v_2\}$ then it
contradicts the fact that $S'$ is a Stackelberg equilibrium since
$v_1$ or $v_2$ could modify their strategy (play $v_3$) and benefit.
Then $T_{v_3} \notin \{v_1,v_2\}$ a short diminishing configuration
as the one of Figure \ref{figcas13} $(f)$ is found.

Suppose that $v_3 \notin Q$.  If $v_5 \in Q$ then $T_{v_5} = v_7$
and  $S'$ is not a Stackelberg equilibrium because $v_3$ could play
$v_1$ instead of $v_5$ and benefit, contradiction. We deduce that
$v_5 \notin Q$. Suppose there is a node $v \notin \{v_3,v_5\}$ such
that $S^*_v=v_7$. That node would be unmatched in $\mathcal{M}^*$
but $(v,v_7,v_5,v_3)$  would be an augmenting path, contradicting
the optimality of $\mathcal{M}^*$. If $S^*_{v_3}=v_7$ then set
$S^*_{v_3}=v_5$. An average diminishing configuration  as the one of
Figure \ref{figcas13} $(g)$ is then found.

\medskip

\noindent Case $B5$: 
Suppose that $v_4 \in Q$. If $T_{v_4} \in \{v_1,v_2\}$ then it
contradicts the fact that $S'$ is a Stackelberg equilibrium since
$v_1$ or $v_2$ could modify their strategy (play $v_4$) and benefit.
Then $T_{v_4} \notin \{v_1,v_2\}$ and a short diminishing
configuration as the one of Figure \ref{figcas13} $(e)$ is found.

Suppose that $v_4 \notin Q$.  If $v_6 \in Q$ then $T_{v_6} = v_7$.
We deduce that $v_4$ is unmatched in $\mathcal{M}'$. Since
$S'_{v_2}=v_4$ and $S'_{v_4}=v_6$, it contradicts the fact that $S'$
is a Stackelberg equilibrium. Therefore $v_6 \notin Q$. Suppose
there is a node $v \neq v_4$ such that $S^*_v=v_7$. That node would
be unmatched in $\mathcal{M}^*$ but $(v,v_7,v_6,v_4)$  would be an
augmenting path, contradicting the optimality of $\mathcal{M}^*$. If
$S^*_{v_4}=v_7$ then set $S^*_{v_4}=v_6$.
 An average diminishing configuration  as the one
of Figure \ref{figcas13} $(h)$ is then found.  \qed

\eprf

Notice that Lemma \ref{car} is obtained with {\em any} maximum matching $\mathcal{M}^*$ compatible with $\langle T,Q\rangle$.

\begin{theorem} \label{correal} One can decide in polynomial time whether a solution $\langle T,Q\rangle$ is feasible. \end{theorem}

\prf Compute a maximum matching $\mathcal{M}^*$ compatible with
$\langle T,Q\rangle$. If $\mathcal{M}^*$ is not optimum in $G$ then
$\langle T,Q\rangle$ is not feasible. From now on suppose that
$\mathcal{M}^*$ is optimum. Let $S^*$ be any Stackelberg equilibrium
compatible with both $\mathcal{M}^*$  and $\langle T,Q\rangle$.
Using Lemma \ref{car}, $\langle T,Q\rangle$ is not feasible iff
$\mathcal{M}^*$ and $\langle T,Q\rangle$ possess a diminishing
configuration. Short and average diminishing configurations contain
a constant number of nodes so we can easily check their existence in
polynomial time.








For every pair of distinct nodes $\{a,b\}$ such that $a$ and $b$ are
matched in $\mathcal{M}^*$ but not together, we check whether a long
diminishing configuration with extremities $a$ and $b$ exists.
Suppose that $a$ (resp. $b$) is
matched with $a'$ (resp. $b'$). 
If there is a node $v \in V \setminus \{a',b'\}$ such that $S^*_v
\in \{a,b\}$ in every Stackelberg equilibrium $S^*$ compatible with
$\langle T,Q \rangle$ then the answer is no. Otherwise every
unmatched neighbor $v$ of $a$ or $b$ can play a strategy $S^*_v
\notin \{a,b\}$ and $S^*$ remains a Stackelberg equilibrium
compatible with $\langle T,Q\rangle$. If $\{a',b'\} \cap Q \neq
\emptyset$ then the answer is also no.  Now consider the graph
$G'=G[V \setminus (Q \cup \{a,b\})]$ to which we add $(a,a')$ and
$(b,b')$. Deciding whether there exists an $a-b$ path in $G'$ which
alternates edges of $\mathcal{M}^*$ and edges not in
$\mathcal{M}^*$, and such that the first and last edge of this path
are respectively $(a,a')$ and $(b,b')$, can be done in $O(n^{2.5})$
steps. This result is due to J. Edmonds and a sketch of proof can be
found in \cite{M95} (Lemma 1.1).  This problem is equivalent to
checking whether a long diminishing configuration with extremities
$a$ and $b$ exists in $G$.\qed

\eprf

We have mentioned in the previous section that for some graphs,
forcing no node is the optimal Stackelberg strategy, leading to an
optimal solution with value $0$. Such a particular case can be
detected in polynomial time by Theorem \ref{correal}. In the
following study of the approximability of the {\sc mfv} problem, we
will focus on instances for which the strategy of at least one node
must be fixed. We will also make the assumption that any vertex has
at most one leaf in his neighborhood. As we will see, this
restriction can be assumed wlog.

\section{Complexity and approximation}

\subsection{The perfect matching case} \label{subsecperfect}

Let ${\cal G}$ be the class of graphs admitting a perfect matching.

\begin{theorem} \label{theorho} For any $\rho\geq 1$, {\sc mfv} restricted to graphs in ${\cal G}$ is $\rho$-approximable in polynomial time if
and only if the minimum {\sc vertex cover} problem (in general
graphs) is $\rho$-approximable in polynomial time.
\end{theorem}

\prf The proof will be done in two steps. In the  first step,
we will give a polynomial time reduction preserving approximation from the minimum {\sc vertex cover} problem to {\sc mfv} restricted to graphs in ${\cal G}$, while in the second step we will produce a polynomial time reduction preserving approximation from {\sc mfv} restricted to graphs in ${\cal G}$ to the minimum {\sc vertex cover} problem.\\

\noindent $\bullet$ First step. Let $G$ be a simple graph, instance
of the minimum {\sc vertex cover} problem. We suppose that
$V(G)=\{v_1, \cdots,v_n\}$ and $E(G)=\{e_1, \cdots, e_m\}$. Let us
build a simple graph $G'$, instance of {\sc mfv}, as follows: take
$G$, add a copy of every vertex and link every vertex to its copy.
More formally we set $V(G')=\{v_1, \cdots,v_n\} \cup \{v'_1,
\cdots,v'_n\}$ and $E(G')=\{e_1, \cdots, e_m\} \cup \{ (v_i,v'_i) :
i \in \{1,\dots,n\} \}$. Remark that $G'$ admits a unique perfect
matching made of all edges $(v_i,v'_i)$. Then $G\in {\cal G}$. We
claim that $G$ admits a vertex cover $C$ of size at most $k$ iff
$G'$ admits a feasible solution of the same size.

\begin{itemize}

\item[$(\Rightarrow)$]    Consider the solution $\langle T , Q
\rangle$ where $Q=C$ and for every $v_i \in C$, set $T_{v_i}=v'_i$.
Since $C$ is a vertex cover, there is no pair of nodes $v_i,v_j \in
V(G) \setminus C$ such that $(v_i,v_j) \in E(G)$. Hence $v_i$ (resp.
$v_j$) can only match with $v'_i$ (resp. $v'_j$).

\medskip

\item[$(\Leftarrow)$] Suppose that there are two nodes $v_i,v_j
\in V(G) \setminus Q$ such that $(v_i,v_j) \in E(G)$. These nodes
can match because they are not forced, contradicting that $\langle
T,Q \rangle$ is a feasible solution ($v_i$ and $v_j$ must match with
$v'_i$ and $v'_j$ respectively). Therefore $Q \cap V(G)$ is a vertex
cover of $G$, of size at most $|Q|$.\\

\end{itemize}

\noindent $\bullet$ Second step. Let $G=(V,E)$ be a graph admitting
a perfect matching, i.e., $G\in {\cal G}$. Let $G'$ be a graph
defined as $V(G')=\{v \in V(G) : d_G(v) >1\}$ and $E(G')=\{(x,y) \in
E(G) : x,y \in V(G')\}$. We claim that there is a vertex cover of
size at most $k$ in $G'$ iff {\sc mfv} has a solution of value at
most $k$ in $G$.

\begin{itemize}

\item[$(\Rightarrow)$]  Let $C$ be a vertex cover of size $k$ in
$G'$. Compute a maximum matching $\mathcal{M}$ of $G$. Build a
solution $\langle T,Q \rangle$ to the {\sc mfv} problem as follows:
force every node of $C$  to follow the matching $\mathcal{M}$. The
matching being perfect, it is always possible. It is clear that $k$
nodes are forced. 

Let us prove that every Stackelberg equilibrium $S$ compatible with
$\langle T,Q \rangle$ induces the optimal matching $\mathcal{M}$.
Take an edge $(u,v)  \in \mathcal{M}$. If both $u$ and $v$ are
forced then $T_u=v$ and $T_v=u$, by construction. Suppose that only
$u$ is forced. We have $T_u=v$ and there is no node $w \neq u$ such
that $T_w=v$, by construction. If there is an unforced node $w \in
\mathcal{N}_G(v)$ then either $w\in V(G')$, it contradicts the fact
that $C$ is a valid vertex cover of $G'$, or $v\in V(G)\setminus
V(G')$, contradicts the fact that $\mathcal{M}$ is a perfect
matching of $G$. Now suppose that neither $u$ nor $v$ is forced. At
least one of them, say $u$, has degree $1$ since otherwise $C$ is
not a valid vertex cover. As previously an unforced node $w \in
\mathcal{N}_G(v)$ would contradict that $C$ is a valid vertex cover.
Then $u$ can only play $v$ and $v$'s rational behavior is to play
$u$.

\medskip

\item[$(\Leftarrow)$]  Take a solution $\langle T,Q \rangle$ with
$|Q|=k$ and build a vertex cover $C:=V(G') \cap Q$. It is clear that
$C$ has size at most $k$. We can observe that $C$ is not a vertex
cover in $G'$ iff there exists an edge $(u,v)$ with $d_G(u)>1$,
$d_G(v)>1$ and $\{u,v\} \cap Q= \emptyset$. Let $\mathcal{M}$ be an
optimal (and perfect) matching induced by a Stackelberg equilibrium
compatible with $\langle T,Q \rangle$. If $u$ and $v$ are matched in
$\mathcal{M}$ then $u$ (resp. $v$) has a matched neighbor $u'\neq v$
(resp. $v' \neq u$). If $u$ (resp. $v$) plays $u'$  (resp. $v'$)
then we get an equilibrium which contradicts the fact that $\langle
T,Q \rangle$ is a feasible solution. If $u$ and $v$ are not matched
in $\mathcal{M}$ then suppose that $u$ is matched with $u'$ while
$v$ is matched with $v'$ (the matching is perfect). If we remove
$(u,u')$, $(v,v')$ and add $(u,v)$ then the state is an equilibrium
(neither $u'$ nor $v'$ can deviate and be matched with a node since
$\mathcal{M}$ is perfect) but the resulting matching is not optimal.
\qed

\end{itemize}
\eprf

The following corollary is based on known results for {\sc vertex
cover} \cite{AK00,Hoc96}.

\begin{corollary}
The {\sc mfv} problem is APX-hard and $2$-approximable in ${\cal G}$.
\end{corollary}

\subsection{General graphs} \label{subsecgeneral}

We are going to describe and analyse an approximation algorithm called \texttt{APPROX}.
It computes a particular maximum matching  $M$ 
and forces some nodes to follow it. The analysis is done in two phases: $(i)$ \texttt{APPROX} gives a $2$-approximation of solutions inducing $M$ (Theorem \ref{partoneofmainresult}), $(ii)$ the worst case ratio between an optimal solution inducing $M$ and a global optimum for the {\sc mfv} problem is $3$ (Theorem \ref{theoMatchingSpecific}). Combining Theorems \ref{partoneofmainresult} and \ref{theoMatchingSpecific} leads to a $6$-approximation.

\medskip

\begin{figure}
\begin{center}
\psfrag{a}{$u$} \psfrag{b}{$v$} \psfrag{c}{$u_1$}
\psfrag{d}{$v_{1}$} \psfrag{e}{$w$} \psfrag{f}{$u_1$}
\psfrag{g}{$v_1$} \psfrag{h}{$u_2$} \psfrag{i}{$v_2$}
\includegraphics[width=65mm]{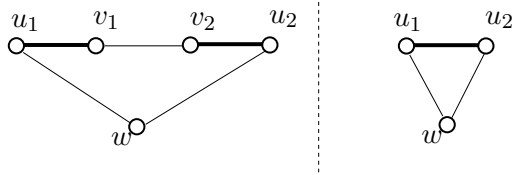}
\caption{Bold edges are in $M$. $w$ has degree 2.} \label{transfo}
\end{center}
\end{figure}

\texttt{APPROX} is as follows. If $G$ is a triangle or a cycle of length 5, then we do not need to
force any vertex. Otherwise, let us consider a maximum matching $M$
such that every leaf in $G$ is matched. We modify the matching $M$
as follows (see Figure~\ref{transfo} for an illustration of the 2
steps):
\begin{enumerate}
    \item\label{step2} While there exists an unmatched vertex $w$ of degree $2$
    such that its neighbors $u_1$ and $u_2$ are matched with $v_1$
    and $v_2$, and $v_1$ and $v_2$ are adjacent: in this cycle of
    length 5 $(w,u_1,v_1,v_2,u_2)$, instead of $(u_1,v_1)$ and $(u_2,v_2)$, take two edges such that the
    unmatched vertex has degree at least 3. This is always
    possible  since $G$ is not a cycle of length 5.
    \item\label{step1} While there exists an unmatched vertex $w$ of degree $2$
    such that its neighbors $u_1$ and $u_2$ are matched together,
    remove $(u_1,u_2)$ from $M$, and add $(u_1,w)$ if $u_1$ has degree 2 or add $(u_2,v)$
    otherwise. Note that the new unmatched vertex has
    degree at least 3 now, since $G$ is not a triangle.
\end{enumerate}

Note that these two steps finish in at most $n$ iterations. At the
end $M$ does not have any of the two ``forbidden'' configurations
(obviously step
\ref{step1} does not create a forbidden configuration of the first type).\\

\begin{figure}
\begin{center}
\psfrag{a}{$u$} \psfrag{b}{$v$} \psfrag{c}{$u_1$}
\psfrag{d}{$v_{1}$} \psfrag{e}{$w$} \psfrag{f}{$u_1$}
\psfrag{g}{$v_1$} \psfrag{h}{$u_2$} \psfrag{i}{$v_2$}
\includegraphics[width=110mm]{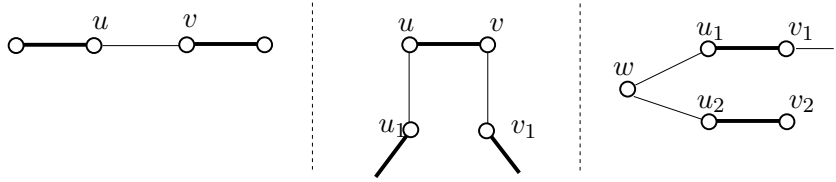}
\caption{Bold edges are in $M$. In case 2, $u_1$ and $v_1$ might be
the same vertex.} \label{figapprox}
\end{center}
\end{figure}

Based on the modified maximum matching $M$, we consider the
following steps  (see Figure~\ref{figapprox} for
an illustration):

\begin{enumerate}
    \item\label{rule1} While there exist in $G$ two unforced adjacent vertices $u,v$ such that $u$ and $v$ are matched in $M$ but not together: force
    $u$ and $v$ according to $M$.
    \item\label{rule2} While there exists in $G$ an edge $(u,v)\in M$ such that
    both $u$ and $v$ are unforced and adjacent to some other matched vertices $u_1$ and $v_1$ (possibly $u_1=v_1$):
    force $u$ and $v$ according to~$M$.
    \item\label{rule3} While there exists an unmatched vertex $w$ which is adjacent to $u_1$ and $u_2$, where $u_1$ is matched with $v_1\neq u_2$,
    $u_2$ is matched with $v_2$, $v_1$ has degree at least 2 and $u_1$, $v_1$ and $u_2$ are not forced: force $u_1$ and $u_2$ according to $M$.
\end{enumerate}

\begin{theorem} \label{partoneofmainresult}
Let $<T^*,Q^*>$ be an optimum solution among the feasible solutions
that are compatible with $M$. Then Algorithm \texttt{APPROX} outputs
a feasible solution $<T,Q>$ such that $|Q|\leq 2|Q^*|$.
\end{theorem}

\prf We first have to show that the output solution is feasible. To
see this, first note that two adjacent vertices $u$ and $v$ that are
matched in $M$ but not together cannot be matched together in an
equilibrium compatible with $<T,Q>$. Indeed, if $u$ and $v$ were not
forced, the edge $(u,v)$ would have been considered in
Step~\ref{rule1} of \texttt{APPROX}. So we only have to show that
for each edge $(u,v)\in M$, at least one vertex between $u$ and $v$
is matched in any equilibrium compatible with $<T,Q>$. If one
vertex, say $u$, is forced, then $v$ has to be matched in any
equilibrium compatible with $<T,Q>$. So let us consider the case
where neither $u$ nor $v$ is forced. Thanks to Step~\ref{rule2} we
know that one vertex, say $u$, is not adjacent to another matched
vertex.
\begin{itemize}\item If $u$ (or $v$) is a leaf, then clearly $v$ (or
$u$) is matched in any equilibrium.
\item Otherwise, we can assume now that $u$ is adjacent to
one or several unmatched vertices $w_1,w_2,\cdots $ such that all
their neighbors but $u$ (and possibly $v$) are forced. Indeed, if
such a $w_i$ were adjacent to an unforced vertex $u_2\neq u,v$,
$w_i$ would have been considered in Step~\ref{rule3}. Then, in an
equilibrium compatible with $<T,Q>$, either $u$ plays $v$ and $v$ is
necessarily matched, or $u$ plays some $w_i$ and in this case $w_i$
is matched either with $u$ or with $v$.
\end{itemize}

Now, we prove that $|Q|\leq 2|Q^*|$. To prove this, we consider a
solution $<T^*,Q^*>$ where no unmatched vertex in $M$ is forced (see Corollary \ref{TheoOptCore} on page \pageref{TheoOptCore}).
We achieve this by showing that at each step of the
algorithm, when we force two vertices we can show that any optimum
solution has to force at least one.

Let us first consider Step~\ref{rule1} of the algorithm. $Q^*$ must contain at least one vertex between $u$ and $v$, otherwise there
exists an equilibrium compatible with $<T^*,Q^*>$ where $u$ and $v$ are
matched, the mate $u'$ of $u$ in $M$ plays $u$, and the mate $v'$ of
$v$ in $M$ plays $v$. This is due to the fact that there is no
unmatched vertex $w$ of degree 2 such that the neighbors of $w$ are
$u'$ and $v'$ (Step~\ref{step2} of the modification of $M$), so each
unmatched vertex can play a vertex different from $u'$ and $v'$.

Now, consider Step~\ref{rule2}.  $Q^*$ must contain at least one
vertex between $u$ and $v$, otherwise there exists an equilibrium
compatible with $<T^*,Q^*>$ where $u$ plays a matched vertex, $v$ plays a
matched vertex. This is due to the fact that there is no unmatched
vertex $w$ of degree 2 such that the neighbors of $w$ are $u$ and
$v$ (Step~\ref{step1} of the modification of $M$), so each unmatched
vertex can play a vertex different from $u$ and $v$.

Now, consider Step~\ref{rule3}. Note that $v_1$ and $v_2$ cannot be
adjacent to an unmatched vertex different from $w$ (an augmenting path would exist), and that at this
step of the algorithm if $v_1$ is adjacent to $v_2$ then $v_2$ is
forced (otherwise $(v_1,v_2)$ would have been considered in
Step~\ref{rule1}). Then $Q^*$ must contain at least one vertex
between $v_1$, $u_1$ and $u_2$, otherwise there exists an
equilibrium compatible with $<T^*,Q^*>$ where $w$ and $u_2$ are matched,
$u_1$ plays $w$, $v_1$ plays another (i.e. different from $u_1$) (matched) vertex and $v_2$
plays $u_2$. This is an equilibrium since as we said if $v_1$ plays
$v_2$ then $v_2$ is forced, and since each unmatched vertex can play
a vertex different from $u_1$ (recall that all the leaves are
matched). To get the final ratio 2, just remark that $v_1$ (and
$u_1$ and $u_2$) will not be considered in another step of the
algorithm (since $v_1$ is not adjacent to an unmatched vertex), so
we do not count twice the same vertex forced in $<T^*,Q^*>$. \qed

\eprf

The next step is the following Theorem.

\begin{theorem}\label{theoMatchingSpecific}
Let $M^*$ be a maximum matching saturating all the leaves of $G$.
Then, there exists a feasible solution $\langle T',Q'\rangle$  of
$G=(V,E)$ and a Stackelberg equilibrium $S'$ compatible with
$\langle T',Q'\rangle$ such that ${\cal M}^{S'}=M^*$ and $|Q'|\leq
3|Q^*|$ where $\langle T^*,Q^*\rangle$ is an optimal solution of
$G$. 
\end{theorem}

For the sake of readability, the proof of Theorem \ref{theoMatchingSpecific}, which requires several intermediate result,
is given later (next section).  Combine Theorems \ref{partoneofmainresult} and \ref{theoMatchingSpecific} to get the main result of this section.

\begin{corollary} \label{mainapproxtheo} \texttt{APPROX} is a $6$-approximation algorithm for the {\sc mfv} problem in general graphs.
\end{corollary}

\section{A proof of Theorem \ref{theoMatchingSpecific}} \label{sectech}


We first see some conditions to build feasible or optimum solutions.
 We will show in particular some interesting properties
that are verified by at least one optimum solution. These properties
will then be used in order to show
Theorem~\ref{theoMatchingSpecific}, a stepping stone of our
approximation result.

Let us first introduce the concepts of {\em basis} of a solution and
of {\it good solution}.

\begin{definition}\label{DefCore} The {\em basis} of a feasible solution $\langle T,Q \rangle$ of $G=(V,E)$
is the edge set $M_b(\langle T,Q \rangle)=\{(u,T_u):u\in Q\}$. A
feasible solution $\langle T,Q \rangle$ of a graph $G=(V,E)$ is
called {\em good} if its basis is a matching of $G$.
\end{definition}

In order to keep simple notations, we will write $M_b$ when no
confusion is possible. The notion of basis of a feasible solution
$\langle T,Q \rangle$ is important. We will show that any optimum
solution is good, and that every maximum matching containing its
basis $M_b$ is induced by a Stackelberg equilibrium $S$ compatible
with $\langle T,Q \rangle$.

Before showing this, let us begin with the following lemma.


%

\begin{lemma} \label{lemMatchingChoosed}
Let $\langle T,Q \rangle$ be a solution of $G=(V,E)$. Let $M$ be a
matching such that $T_i=j$ each time $(i,j) \in M$ and $i \in Q$. There exists a
Stackelberg equilibrium $S$ compatible with $\langle T,Q \rangle$ such that
$M\subseteq {\cal M}^S$. In particular, if $M$ is maximum, then
$M={\cal M}^S$.
\end{lemma}

\prf Consider a matching $M'$ containing $M$, compatible with
$\langle T,Q \rangle$, and maximal w.r.t. these properties. This
means that if $u$ is not matched, then either it is forced, or all
its neighbors are either matched or forced not toward $u$. Let $S$
be a state such that if $(u,v)$ in $M'$ then $u$ plays $v$ and $v$
plays $u$, otherwise if $u$ is unmatched in $M'$ then it plays one
of its neighbors ($T_u$ if $u$ is forced). $S$ is a Stackelberg
equilibrium compatible with $M'$, hence with $M'={\cal M}^S$ while
$M \subseteq M'$. \qed \eprf

As a consequence of Lemma~\ref{lemMatchingChoosed}, the basis $M_b$
of a good solution is contained in some maximum matching of the
graph. Moreover, for any maximum matching $M^*$ which contains
$M_b$, there is a Stackelberg equilibrium $S$ compatible with
$\langle T,Q \rangle$ such that ${\cal M}^{S}=M^*$.

Now, we prove that every optimum solution is good, via the following
theorem.

\begin{theorem}\label{TheoGood}
For any feasible solution $\langle T,Q \rangle$ of $G=(V,E)$, if it
is not good then we can  find in polynomial time a good feasible
solution $\langle T^0,Q^0 \rangle$ of $G=(V,E)$ such that
$Q^0\subset Q$ and $T^0$ is the restriction of $T$ to the vertices
in $Q^0$.
\end{theorem}

\begin{figure}
\begin{center}
\psfrag{a}{$u$} \psfrag{b}{$v$} \psfrag{c}{$w$}
\includegraphics[width=60mm]{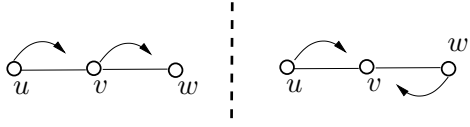}
\caption{The two cases} \label{IoBaby}
\end{center}
\end{figure}

\prf Let $\langle T,Q \rangle$ be a feasible solution of $G=(V,E)$
which is not good, i.e. the basis $M_b=\{(u,T_u):u\in Q\}$ of
$\langle T,Q \rangle$
 is not a matching of $G=(V,E)$. So, there
are $u,v\in Q$ such that either $T_u=v$ and $T_v=w\neq u$, or
$T_u=T_w=v$ where $v\neq u,w$ (see Figure~\ref{IoBaby}). In Case~2,
$v$ is not forced (otherwise it is Case~1).

In both cases, we will show that $\langle T\setminus
\{T_u\},Q\setminus \{u\} \rangle$ is feasible. For this, let $M^*$
be a maximum matching compatible with $\langle T,Q \rangle$
containing $(v,w)$  - this is possible in Case 1 since $w$ cannot be
forced to a vertex different from $v$ (otherwise $M^*$ is not
maximum), and in Case 2 since $v$ is not forced. Of course, $M^*$ is
compatible with $\langle T\setminus \{T_u\},Q\setminus \{u\}
\rangle$. If $\langle T\setminus \{T_u\},Q\setminus \{u\} \rangle$
is not feasible then, using Lemma \ref{car}, we know that $\langle
T\setminus \{T_u\},Q\setminus \{u\} \rangle$ and $M^*$ possess a
diminishing configuration based on a Stackelberg equilibrium $S$. In
$S$, we can assume that $u$ plays $v$ since $u$ is unmatched in
$M^*$ and $v$ is matched. Then $S$ is also an equilibrium compatible
with $\langle T,Q \rangle$. Since $\langle T,Q\rangle$ is feasible,
the diminishing configuration must contain $u$ as an unforced
vertex. Since $u$ is unmatched, the only configurations are $(g)$
and $(h)$ with $x=u$. In both cases, $v$ or $w$ being forced, $v\neq
v_1,v_2$ and $w\neq v_1,v_2$. If there is a $(g)$ configuration
w.r.t. $\langle T\setminus \{T_u\},Q\setminus \{u\} \rangle$ and
$M^*$, then there is an $(f)$ configuration w.r.t. $\langle T,Q
\rangle$ and $M^*$, composed of vertices $v_1,v_2,u$ and $v$.
Alternatively, if there is an $(h)$ configuration w.r.t. $\langle
T\setminus \{T_u\},Q\setminus \{u\} \rangle$ and $M^*$, then there
is an $(e)$ configuration w.r.t. $\langle T,Q \rangle$ and $M^*$.
Contradiction.

Hence, by applying the previous process, we obtain in linear time a
feasible solution $\langle T^0,Q^0 \rangle$ of $G=(V,E)$ such that
$T^0\subset T$ and $Q^0\subset Q$. By construction, the basis of
$\langle T^0,Q^0 \rangle$, $M_b=\{(u,T^0_u):u\in Q^0\}$ is a
matching. \qed \eprf

Using Theorem \ref{TheoGood}, we easily deduce the following result:

\begin{corollary}\label{CorGood}
Any minimal (for inclusion) feasible solution is good. In
particular, any optimal feasible solution is good.
\end{corollary}

Now, we show that, informally, we can suppose that in some optimum
solution the forced vertices are matched in a given matching.

\begin{figure}
\begin{center}
\psfrag{a}{$x_0$} \psfrag{b}{$x_1$} \psfrag{c}{$x_2$}
\includegraphics[width=50mm]{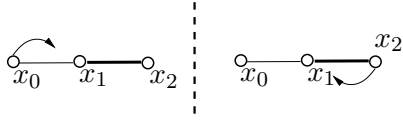}
\caption{$\langle T,Q \rangle$ appears on the left side and $\langle T^0,Q^0
\rangle$ appears on the right side. In dark line an edge of ${\cal M}^S$.} \label{lemmaUselessGoodfig1}
\end{center}
\end{figure}

\begin{lemma}\label{lemmaUselessGood}
Let $\langle T,Q \rangle$ be a good feasible solution of $G=(V,E)$.
If there exists a Stackelberg equilibrium $S$ compatible with
$\langle T,Q \rangle$ and a node $x_0\in Q$ unmatched in ${\cal
M}^S$, then $\langle T^0,Q^0 \rangle$ is a good feasible solution of
$G=(V,E)$ where $Q^0=Q\cup\{x_2\}\setminus \{x_0\}$,
$x_2=S_{T_{x_0}}$ and $T^0_{x_2}=T_{x_0}$ (see Figure
\ref{lemmaUselessGoodfig1} for an illustration). Moreover, $S$ is
compatible with $\langle T^0,Q^0 \rangle$.
\end{lemma}

\prf Let  $\langle T,Q \rangle$ be a good feasible solution of
$G=(V,E)$ and  $S$ be an equilibrium compatible with $\langle T,Q
\rangle$ verifying the hypothesis. Let $x_1=T_{x_0}$. ${\cal M}^S$
is a maximum matching where $x_0$ is unmatched. ${\cal M}^S$ is also
compatible with $\langle T^0,Q^0 \rangle$ (since $(x_1,x_2)\in {\cal
M}^S$) (see Figure \ref{lemmaUselessGoodfig1}). Suppose that
$\langle T^0,Q^0 \rangle$  is not a feasible solution. By Lemma
\ref{car}, $\langle T^0,Q^0 \rangle$ and ${\cal M}^S$ possess a
diminishing configuration on the base of a Stackelberg equilibrium
$S'$.
Note that since $x_1$ is matched and $x_0$ is not matched in ${\cal
M}^S$, we can assume that $S'_{x_0}=x_1$. Then, it is easy to see
that $S'$ is also a Stackelberg equilibrium compatible with $\langle
T,Q\rangle$. Since $\langle T,Q\rangle$ is feasible, it means that
the diminishing configuration w.r.t. $\langle T^0,Q^0\rangle$ and
${\cal M}^S$ contains $x_2$ as a necessarily forced vertex and/or
$x_0$ as a necessarily unforced vertex. But $x_2$ is matched and
there is no matched and necessarily forced vertex in the diminishing
configurations. Then, it is either configuration $(g)$ or $(h)$,
where $x=x_0$. Since $S'_{x_0}=x_1$ and $x_1$ is matched with $x_2$
which is forced in $\langle T^0,Q^0\rangle$, $x_1\neq v_1,v_2$.
Then, a $(g)$ configuration w.r.t. $\langle T^0,Q^0\rangle$ and
${\cal M}^S$ corresponds to an $(f)$ configuration w.r.t. $\langle
T,Q\rangle$ and ${\cal M}^S$, and an $(h)$ configuration w.r.t.
$\langle T^0,Q^0\rangle$ and ${\cal M}^S$ corresponds to an $(e)$
configuration w.r.t. $\langle T,Q\rangle$ and ${\cal M}^S$.
Contradiction.

It is easy to check that $\langle T^0,Q^0\rangle$ is a good solution
(no vertex $x_3$ is forced toward $x_2$ in $\langle T,Q\rangle$
otherwise an augmenting path $(x_3,x_2,x_1,x_0)$ exists, meaning
that
${\cal M}^S$ is not maximum). 
\qed \eprf

From Lemma~\ref{lemmaUselessGood}, we deduce the following
corollary.

\begin{corollary}\label{TheoOptCore}
Let $\langle T,Q\rangle$ be a feasible (resp. optimum) solution and
$S$ be a Stackelberg equilibrium compatible with $\langle T,Q\rangle$. There exists a
feasible (resp. optimum) solution $\langle T^0,Q^0\rangle$
compatible with $S$ such that every forced vertex (in $Q^0$) is
matched in ${\cal M}^S$. In particular, ${\cal M}^S$ contains the
basis of $\langle T^0,Q^0\rangle$.
\end{corollary}

%

Given a graph $G=(V,E)$, we denote by $L$ the set of leaves of
$G$ and for $\ell\in L$, $w_\ell$ is the neighbor of $\ell$.
Finally, $M_L=\{(\ell,w_\ell):\ell\in L\}$. Note that if several
leaves are adjacent to the same vertex, then the problem is
obviously equivalent when we remove in the graph all these leaves
but one. {\it In the sequel we will assume that two leaves are never
adjacent to the same vertex}. In other words, $M_L$ is a matching.

In the following lemma, we mainly prove that there is an optimal
solution $\langle T,Q \rangle$ and a compatible Stackelberg equilibrium
whose induced matching contains $M_L$.

\begin{lemma}\label{lemmaLeafs}
From any good feasible solution $\langle T,Q \rangle$  of $G=(V,E)$,
one can  find in polynomial time a good feasible solution $\langle T^0,Q^0
\rangle$  such that $|Q_0|\leq |Q|$ and there exists a Stackelberg equilibrium $S_0$ compatible with $\langle T^0,Q^0 \rangle$
satisfying $M_L\subseteq {\cal M}^{S^0}$.
\end{lemma}

\begin{figure}
\begin{center}
\psfrag{a}{$\ell$} \psfrag{b}{$w_\ell$}
\psfrag{c}{$x$}\psfrag{d}{$A$} \psfrag{e}{$T_{w_\ell}$}
\includegraphics[width=60mm]{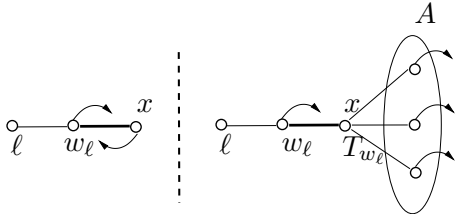}
\caption{Left: case $(1)$. Right: case $(2)$.}
\label{lemmaUseleavesfig}
\end{center}
\end{figure}

\prf For the sake of contradiction, suppose that  $\langle
T,Q\rangle$ is such that for every equilibrium $S$ compatible with
$\langle T,Q\rangle$, $M_L\nsubseteq {\cal M}^{S}$. Hence,  we must
have $w_\ell\in Q$ and $x=T_{w_\ell}\neq \ell$ for some $\ell\in L$
(note that $\ell\notin Q$ by Theorem \ref{TheoGood}). We will prove
that two alternatives are possible: either case $(1)$ $x\in Q\cap
{\cal N}_G(w_\ell)$ and $T_x=w_\ell$, or case $(2)$ $A\subseteq Q$
and $\forall v\in A$, $T_v\neq T_{w_\ell}$ where $A={\cal
N}_G(x)\setminus \{w_\ell\}$. Remark that since $\langle T,Q\rangle$
is a good feasible solution, if $x\in Q\cap {\cal N}_G(w_\ell)$,
then $T_x=w_\ell$ (and such $x$ is unique) and if $A\subseteq Q$,
then $\forall v\in A$, $T_v\neq T_{w_\ell}$ (see Figure
\ref{lemmaUseleavesfig} for an illustration of the two cases).

Assume that cases $(1)$ and $(2)$ do not hold. Consider the partial
state $S'$ with $S'_x=y$ and $S'_y=x$ where $y\in N_G(x)\setminus
Q$. Obviously, $x\notin L$ ($M_L$ is a matching) and such a $y$
exists since $(1)$ and $(2)$ do not hold. We can extend $S'$ into a
forced Nash equilibrium $S^0$ compatible with $\langle T,Q\rangle$.
We get that ${\cal M}^{S^0}\cup \{(\ell,w_\ell)\}$ is a matching,
contradiction.

\begin{itemize}

\item[$\bullet$] Assume that Case $(1)$ happens. Consider $\langle T^0,Q^0\rangle$
which is the same as $\langle T,Q\rangle$ up to the facts that $x$
is unforced and $T^0_{w_\ell}=\ell$. We can check that it is a good
feasible solution.

\item[$\bullet$] Assume that Case $(2)$ holds. In this case, consider $\langle T^0,Q^0\rangle$
which is the same as $\langle T,Q\rangle$ up to the fact that
$T^0_{w_\ell}=\ell$. As previously, we can check that $\langle
T^0,Q^0\rangle$ is a good feasible solution.

\end{itemize}

By repeating this procedure for every  $\ell\in L$ such that
$w_\ell\in Q$  and $T_{w_\ell}\neq \ell$, we obtain the expected
result. \qed \eprf

Based on the previous intermediate results, we are now able to give
a proof of Theorem~\ref{theoMatchingSpecific} (we reproduce its
statement for the sake of readability).

\medskip

\noindent {\bf Theorem \ref{theoMatchingSpecific}.} {\em
Let $M^*$ be a maximum matching saturating all the leaves of $G$.
Then, there exists a feasible solution $\langle T',Q'\rangle$  of
$G=(V,E)$ and a Stackelberg equilibrium $S'$ compatible with
$\langle T',Q'\rangle$ such that ${\cal M}^{S'}=M^*$ and $|Q'|\leq
3|Q^*|$ where $\langle T^*,Q^*\rangle$ is an optimal solution of
$G$. 
}

\prf Let $G=(V,E)$ be a graph such that $M_L$ is a matching. Let
$M^*$ be a maximum matching saturating all the leaves of $G=(V,E)$.
Consider an optimal solution $\langle T^*,Q^*\rangle$ of $G$ such
that there is a Stackelberg equilibrium $S$ where $|{\cal M}^{S}\cap
M^*|$ is as large as possible. Moreover, by Corollary
\ref{TheoOptCore}, we assume that the basis of $\langle
T^*,Q^*\rangle$ is included in ${\cal M}^{S}$; in particular, all
vertices of $Q^*$ are matched in ${\cal M}^{S}$. Also, assume that
$S$ satisfies Lemma \ref{lemmaLeafs}, i.e., $M_L\subseteq {\cal
M}^{S}\cap M^*$. It it was not the case, $|{\cal M}^{S}\cap M^*|$
would not be maximum (the transformation in the proof of the Lemma
would increase $|{\cal M}^{S}\cap M^*|$).

Suppose that ${\cal M}^{S}\neq M^*$ (otherwise, we are done). Thus, there
exists some alternating path or alternating cycle of ${\cal
M}^{S}\Delta M^*$ where $\Delta$ is the symmetric difference
operator. The solution  $\langle T',Q'\rangle$ will be built by
first examining the partial graph induced by ${\cal M}^{S}\Delta
M^*$ and then the partial graph induced by ${\cal M}^{S}\cap M^*$.

Let $G'=(V,{\cal M}^{S}\Delta M^*)$ and consider a connected
component $G_1=(V_1,E_1)$ of $G'$ with $|V_1|\geq 2$ (then,
$|E_1|\geq 2$ since $M$ and ${\cal M}^{S}$ are maximum matchings).
The following property holds:

\begin{property}\label{property1MatchingSpecific}
If $G_1$ is a cycle then $2|Q^*\cap V_1|\geq |V_1|$. If it is a path then
$2|Q^*\cap V_1|\geq |V_1|-1$
\end{property}

\prf We study two cases: $G_1$ is an even cycle or an even path.
Note that in both cases, if the inequality is false then there is in
$G_i$ an edge in
$M^*$ with no extremity in $Q^*$.\\

Assume  $G_1$ is an even cycle $(x_1,x_2,\cdots,x_{2p},x_1)$, where
edges $(x_{2i-1},x_{2i})$ are in $M^*$. Wlog. let say that
$(x_{3},x_{4})$ is an edge such that $x_{3},x_{4}\not \in Q^*$.

Suppose that $x_2$ is in $Q^*$. Then let the state $S'$ be the same
state as $S$ up to the facts that $x_3$ plays $x_4$, $x_4$ plays
$x_3$, and all the (unmatched) neighbors of $x_5$ do not play $x_5$
(but another (matched) vertex). This is possible since $x_5$ is not
adjacent to a leaf, because unmatched vertices (in ${\cal M}^S$) are
not in $Q^*$ and the unmatched vertices form an independent set
(${\cal M}^S$ is a maximum matching). $S'$ is an equilibrium
compatible with $\langle T^*,Q^*\rangle$ with ${\cal M}^{S'}$ not
maximum, contradiction.

So neither $x_2$ nor $x_5$ are in $Q^*$. If $x_{2}$ or $x_{5}$ has
an unmatched neighbor $u$, say $x_{5}$, then there is an equilibrium
$S'$ compatible with $\langle T^*,Q^*\rangle$ with $|{\cal
M}^{S'}\cap M^*|>|{\cal M}^{S}\cap M^*|$: $S'$ is the same as $S$ up
to the facts that $x_{5}$ and $u$ are matched (play each other),
$x_{3}$ and $x_{4}$ are matched, and no vertex plays $x_{2}$.

Finally, if neither $x_{2}$ nor $x_{5}$ have an unmatched neighbor,
then consider the state $S'$ which is the same as $S$ up to the
facts that $x_{3}$ plays $x_{4}$ and $x_{4}$ plays $x_3$. $S'$ is
compatible with $\langle T^*,Q^*\rangle$ but ${\cal M}^{S'}$ is not
maximum, contradiction.\\

Now, suppose that $G_1=(V_1,E_1)$ is an even path
$(x_1,\cdots,x_{2p+1})$. Let $(x_{2i-1},x_{2i})$ be an edge in $M^*$
with no extremity in $Q^*$. The previous arguments work unless the
edge under consideration is $(x_1,x_2)$. But then the state $S'$
which is the same as $S$ up to the facts that $x_1$ plays $x_2$,
$x_2$ plays $x_1$ and no vertex plays $x_3$ is a Stackelberg
equilibrium  compatible with $\langle T^*,Q^*\rangle$ such that
$|{\cal M}^{S'}\cap M^*|>|{\cal M}^{S}\cap M^*|$, contradiction.\qed

\eprf

Now, let  $x_1$ be a vertex unmatched by ${\cal M}^{S}$ and matched
by $M^*$ ($x_1$ is a leaf of $G_1$ where $G_1$ is a path of length
at least 2 of $G'=(V,{\cal M}^{S}\Delta M^*)$). We are interested in
edges $e_i=(u_i,v_i)\in{\cal M}^{S}\cap M^*$ with $i\leq q$ such
that $(x_1,v_i)\in E$.

\begin{property}\label{property2MatchingSpecific}
Among the edges of $\{e_1,\dots,e_q\}$ whose extremities are not
leaves, there is at most one edge, say $e_1=(u_1,v_1)$, such that
$u_1,v_1\notin Q^*$.
\end{property}

\begin{figure}
\begin{center}
\psfrag{a}{$x_1$} \psfrag{b}{$x_2$} \psfrag{c}{$x_3$}
\psfrag{d}{$v_1$} \psfrag{e}{$u_1$} \psfrag{f}{$v_2$}
\psfrag{g}{$u_2$} \psfrag{h}{$w$} \psfrag{i}{$z$}
\includegraphics[width=45mm]{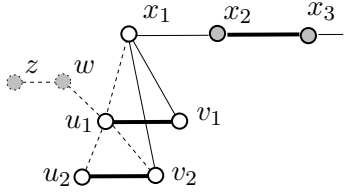}
\caption{Illustration of Property \ref{property2MatchingSpecific}.}
\label{illusprop2}
\end{center}
\end{figure}

\prf For the sake of contradiction, assume that there is another
edge $e_2=(u_2,v_2)\in{\cal M}^{S}\cap M^*$ such that $(x_1,v_2)\in
E$, $u_2,v_2\notin Q^*$ and $u_2,v_2$ are not leaves. See Figure
\ref{illusprop2}. We show that $\langle T^*,Q^*\rangle$ is not
feasible.

We study several cases depending on the neighborhood of $u_1$. 
Remark that neither $u_1$ nor $u_2$ is adjacent to a vertex
unmatched in ${\cal M}^S$ and different from $x_1$, otherwise an
augmenting path exists. In every case, we will assume that if a
vertex $w$ unmatched  in ${\cal M}^{S}$ is adjacent to $v_1$, then
$S_w\neq v_1$  because $w$ is not a leaf of $G$. As previously
indicated it is always possible.

\begin{itemize}

\item[$\bullet$] $(u_1,x_1)\in E$. In this case we have a diminishing configuration $(g)$
(recall that $x_1\notin Q^*$ by Lemma \ref{lemmaUselessGood}) w.r.t.
$\langle T^*,Q^*\rangle$ and ${\cal M}^S$.

\item[$\bullet$] $(u_1,v_2)\in E$. We get a diminishing configuration $(a)$ on
the path $(v_1,u_1,v_2,u_2)$.

\item[$\bullet$] $(u_1,u_2)\in E$. $v_1$ and $v_2$ cannot be adjacent to an unmatched
vertex different from $x_1$, otherwise an augmenting path exists.
Then there is a diminishing configuration $(a)$ on the path
$(v_1,u_1,u_2,v_2)$.

\item[$\bullet$] Finally, since $u_1$ is not a leaf, it is adjacent to a vertex $w\neq u_2,v_2,x_1$ which is
necessarily matched with $z$ in ${\cal M}^S$. We get a diminishing
configuration $(h)$ on the path $(z,w,u_1,v_1,x_1,v_2,u_2)$.\qed
\end{itemize}

\eprf

Assume that there are $t$ connected components $G_1,\dots,G_t$ in
the graph $G'=(V,{\cal M}^{S}\Delta M^*)$, and that in $G_i$, $p_i$
vertices are in $Q^*$. Let $F$ be the edges $e_i=(u_i,v_i)\in
M^*\cap {\cal M}^S$ such that $u_i,v_i\notin Q^*$, $u_i$ is not a
leaf and $v_i$ is adjacent to a vertex unmatched by ${\cal M}^{S}$
and matched by $M^*$ (as indicated in Property
\ref{property2MatchingSpecific}), let $R$ be the set of edges in
$M^*\cap {\cal M}^S$ with at least one extremity in $Q^*$, and let
$q=|F|$ and $r=|R|$.

We are ready to build the solution $\langle T',Q'\rangle$ compatible
with $M^*$ as follows. For each edge $(x,y)\in M^*$: if it is in
some $G_i$ or if it is in $R$, then we force both $x$ and $y$
(toward each other). Otherwise, if it is in $F$ then one of its
extremities, say $y$, is adjacent to a vertex matched by $M^*$ but
not by ${\cal M}^S$: we force $y$ (toward $x$). The other vertices
are unforced (in particular some edge $(u,v)\not\in F$ with $u$ and
$v$ not in $Q^*$ may exist; we do not force any of its extremity).
Note that $Q^*\subseteq Q'$.

By definition, we get that:

\begin{equation}\label{eq1theoMatchingSpecific}
|Q^*|\geq \sum_{i=1}^t p_i+r
\end{equation}

By construction of $\langle T',Q'\rangle$ and Property
\ref{property1MatchingSpecific} we get that:

\begin{equation}\label{eq2theoMatchingSpecific}
|Q'|\leq 2\sum_{i=1}^t p_i+2r+q\leq 2|Q^*|+q
\end{equation}

Finally, using Property \ref{property1MatchingSpecific}, we get
that $\forall i\leq t$, $p_i\geq 1$. Thus, by Property
\ref{property2MatchingSpecific} we obtain:

\begin{equation}\label{eq3theoMatchingSpecific}
q\leq t\leq \sum_{i=1}^t p_i \leq |Q^*|
\end{equation}

Using inequalities (\ref{eq1theoMatchingSpecific}),
(\ref{eq2theoMatchingSpecific}) and (\ref{eq3theoMatchingSpecific}),
we obtain $|Q'|\leq 3|Q^*|$.

By construction $M^*$ is compatible with
$\langle T',Q'\rangle$. Let us finally prove that $\langle T',Q'\rangle$ is a
feasible solution. For the sake of contradiction, suppose that there
is a Stackelberg equilibrium $S^2$ compatible with  $\langle T',Q'\rangle$ and such
that $|{\cal M}^{S^2}|< |M^*|$.

Let $G'=(V,{\cal M}^{S^2}\Delta M^*)$. $G'$ is made of even cycles,
even paths and odd paths where the two final edges are in $M^*$.
Moreover, such an odd path must exist in $G'$. Denote it by $P=(v_1,\dots,v_{2l})$. Note that every vertex of $G_i$ matched in $M^*$ is
forced in $\langle T',Q'\rangle$, so each edge of $P$ which is in $M^*$ is also
in ${\cal M}^S$.

Suppose that there is a vertex $w$ unmatched in ${\cal M}^S$ with
${\cal N}(w)=\{v_1,v_{2l}\}$. Then $w$ plays $v_1$ or $v_{2l}$ in
$S^2$, and since $S^2$ is an equilibrium, $v_1$ (or $v_{2l}$) is in
$Q'$. In particular, $v_2$ (or $v_{2l-1}$) is matched in $S^2$,
meaning that $P$ contains at least 2 edges. Note that vertices
$v_2,\cdots,v_{2l-1}$ are not in $Q'$ hence not in $Q^*$. If $v_1$
(or $v_{2l}$) is in $Q^*$, then we get a diminishing configuration
$(a)$ w.r.t. $\langle T^*,Q^*\rangle$, contradiction. But otherwise
$v_1$ (or $v_{2l})$ is in $Q'\setminus Q^*$ and $v_2\not\in Q'$,
meaning that $v_1$ is adjacent to a vertex $x_1$ matched in $M^*$
but not in ${\cal M}^S$, and there would be an augmenting path in
${\cal M}^S$.

So we suppose now that there is no vertex $w$ unmatched in ${\cal
M}^S$ with ${\cal N}(w)=\{v_1,v_{2l}\}$. Then if $P$ has at least 2
edges, there exists a diminishing configuration $(a)$ w.r.t.
$\langle T^*,Q^*\rangle$. But if $P$ is $(v_1,v_2)$, $v_1$ and $v_2$
are not in $Q'$, they are not leaves (otherwise one of them would be
matched in ${\cal M}^{S^2}$). Let $t$ and $x$ be the vertices played
by $v_1$ and $v_2$ in $S^2$ (possibly $x=t$). If $t$ and $x$ are
matched in ${\cal M}^S$ then there is a diminishing configuration
$(b)$, $(c)$ or $(d)$ w.r.t. $\langle T^*,Q^* \rangle$. If $t$ and
$x$ are unmatched in ${\cal M}^S$, then $x=t$ (or there is an
augmenting path), $x$ is adjacent to a third vertex $y$ which is
necessarily matched in ${\cal M}^S$, and we get a diminishing
configuration $(g)$ w.r.t. $\langle T^*,Q^*\rangle$. Finally, $x$ is
unmatched and $t$ is matched in ${\cal M}^S$. $x$ is adjacent to a
vertex $y\neq v_1,v_2$ which is necessarily matched in ${\cal
M}^{S}$. We get a diminishing configuration $(h)$ on vertices
$(S_t,t,v_1,v_2,x,y,S_y)$, contradiction.


In conclusion, $\langle T',Q'\rangle$ is a feasible solution and the
proof is complete. \qed \eprf

\section{Conclusion} \label{secconcl}


The $6$-approximation algorithm for {\sc mfv} in general graphs is
achieved in two steps, namely a $2$-approximation and a
$3$-approximation. One can show that the analysis of both steps is
(asymptotically) tight; an interesting future work would be to reach
a better approximation algorithm by considering a global approach.
Due to the importance of the assignment problem, it is also worth
studying the complexity and approximability of {\sc mfv} in
bipartite graphs.

The model studied for the {\sc mvf} problem can be extended in many directions. For example,
require a feasible not to reach a social optimum in any case but to reach an approximation of it.
In this paper we considered a fixed unit cost for every forced node but it is natural to study
a version where the nodes have different cost for being forced.

Finally, our study focuses on Nash equilibria, i.e. states resilient
to deviations by any single player. An interesting extension is to
deal with simultaneous deviations by several players. In particular,
simultaneous deviations of two players is considered in the stable
marriage problem mentioned in introduction, where a solution is
stable if there is no pair $(x,y)$ of players where both $x$ and $y$
would be happier to be together than with their respective
husband/wife in the solution. In our setting, a state is called a
$2$-strong equilibrium if it is resilient to deviations of at most 2
players. More generally, a state is a {\it $k$-strong equilibrium}
if it is resilient to deviations of at most $k$-players, and it is a
{\it strong equilibrium} if this is true for coalitions of arbitrary
size. Then the $k$-strong (resp. strong) price of anarchy is defined
as the price of anarchy but for $k$-strong (resp. strong)
equilibria.

Dealing with this last issue, we show that the notions of
$2$-strong, $k$-strong and strong equilibria coincide for the game
we consider, and that states resilient to deviations of several
players are much better in term of social welfare than simple Nash
equilibria.
\begin{proposition} A $2$-strong equilibrium is a strong equilibrium.
\end{proposition}

\prf Let $S$ be a $2$-strong equilibrium and suppose that $S$ is not
a strong equilibrium. There exists a coalition of players $C$, with
$|C|>2$, whose members can deviate from $S$ and benefit. For every
member of $C$, the utility is equal to $0$ before the deviation, and
then equal to $1$. This means that $S$ contains some pairs of
unmatched neighbors, and it contradicts the fact that  $S$ is a
$2$-strong equilibrium. \qed

\eprf

\begin{theorem} The strong price of anarchy is $1/2$.
\end{theorem}

\prf Let $S$ be a $2$-strong equilibrium whereas $S^*$ is an optimum
state. For every edge $(i,j) \in E$, we have $\max \{u_i(S),u_j(S)\}
\ge 1$. Take a maximum cardinality matching $\mathcal{M}^*$ and use
the previous inequality to get that $\mathcal{SW}(S) = \sum_{i \in
V} u_i(S) \ge \sum_{\{i,j\} \in \mathcal{M}^*} u_i(S) + u_j(S) \ge
\sum_{\{i,j\} \in \mathcal{M}^*} \max \{ u_i(S) , u_j(S) \}.$ It
follows that $\mathcal{SW}(S) \ge
|\mathcal{M}^*|=\mathcal{SW}(S^*)/2.$ Take a path of length $3$ as a
tight example. \qed

\eprf

However, considering the {\sc mfv} problem for strong equilibria is
an interesting topic that is worth being considered in some future
works.

\bibliographystyle{abbrv}
\bibliography{biblio}

\begin{thebibliography}{10}

\bibitem{AK00}
P.~Alimonti and V.~Kann.
\newblock Some apx-completeness results for cubic graphs.
\newblock {\em Theor. Comput. Sci.}, 237(1-2):123--134, 2000.

\bibitem{BHS10}
V.~Bonifaci, T.~Harks, and G.~Sch{\"a}fer.
\newblock Stackelberg routing in arbitrary networks.
\newblock {\em Math. Oper. Res.}, 35(2):330--346, 2010.

\bibitem{CKN09}
G.~Christodoulou, E.~Koutsoupias, and A.~Nanavati.
\newblock Coordination mechanisms.
\newblock {\em Theor. Comput. Sci.}, 410(36):3327--3336, 2009.

\bibitem{GS62}
D.~Gale and L.~S. Shapley.
\newblock College admissions and the stability of marriage.
\newblock {\em The American Mathematical Monthly}, 69(1):9--15, 1962.

\bibitem{Hoc96}
D.~S. Hochbaum.
\newblock {\em Approximation Algorithms for {NP}-Hard Problems.}
\newblock PWS Publishing Company, Boston, MA, 1996.

\bibitem{ILMS05}
N.~Immorlica, L.~Li, V.~S. Mirrokni, and A.~S. Schulz.
\newblock Coordination mechanisms for selfish scheduling.
\newblock In X.~Deng and Y.~Ye, editors, {\em WINE}, volume 3828 of {\em
  Lecture Notes in Computer Science}, pages 55--69. Springer, 2005.

\bibitem{Irving85}
R.~W. Irving.
\newblock An efficient algorithm for the "stable roommates" problem.
\newblock {\em J. Algorithms}, 6(4):577--595, 1985.

\bibitem{IMMM99}
K.~Iwama, D.~Manlove, S.~Miyazaki, and Y.~Morita.
\newblock Stable marriage with incomplete lists and ties.
\newblock In J.~Wiedermann, P.~van Emde~Boas, and M.~Nielsen, editors, {\em
  ICALP}, volume 1644 of {\em Lecture Notes in Computer Science}, pages
  443--452. Springer, 1999.

\bibitem{KP09}
E.~Koutsoupias and C.~H. Papadimitriou.
\newblock Worst-case equilibria.
\newblock {\em Computer Science Review}, 3(2):65--69, 2009.

\bibitem{M95}
Y.~Manoussakis.
\newblock Alternating paths in edge-colored complete graphs.
\newblock {\em Discrete Applied Mathematics}, 56(2-3):297--309, 1995.

\bibitem{RT04}
T.~Roughgarden.
\newblock Stackelberg scheduling strategies.
\newblock {\em SIAM J. Comput.}, 33(2):332--350, 2004.

\bibitem{RT02}
T.~Roughgarden and {\'E}.~Tardos.
\newblock How bad is selfish routing?
\newblock {\em J. ACM}, 49(2):236--259, 2002.

\bibitem{CS07}
C.~Swamy.
\newblock The effectiveness of stackelberg strategies and tolls for network
  congestion games.
\newblock In N.~Bansal, K.~Pruhs, and C.~Stein, editors, {\em SODA}, pages
  1133--1142. SIAM, 2007.

\end{thebibliography}

\end{document}